\documentclass[manuscript,screen]{acmart}

\setcopyright{none}
\renewcommand\footnotetextcopyrightpermission[1]{}
\acmDOI{}
\acmISBN{}

\usepackage{amsmath}
\usepackage{graphicx}
\usepackage{multirow}
\usepackage{xcolor}

\AtBeginDocument{%
  }

\begin{document}

\title{On the Lexical Superstition of Large Language Models for Code Comprehension: Re-evaluation on Code of Low Lexical Quality}

\author{Xin Shen}
\email{shenx@lamda.nju.edu.cn}
\orcid{0009-0008-1158-2073}
\author{Sanzhuo Xi}
\email{xisz@lamda.nju.edu.cn}
\orcid{0009-0007-5098-5855}
\author{Yali Du}
\email{duyl@lamda.nju.edu.cn}
\orcid{0000-0001-7759-3906}
\author{Ming Li}
\authornote{Corresponding author.}
\email{lim@lamda.nju.edu.cn}
\orcid{0000-0001-7977-5500}
\affiliation{%
  \institution{Nanjing University}
  \city{Nanjing}
  \country{China}
}

\renewcommand{\shortauthors}{Shen et al.}

\begin{abstract}
Recent advances in large language models (LLMs) have made them widely used for code-related tasks. Identifier names are statistically informative in naturally occurring code, but their information is not always reliable. We investigate whether current LLMs assign disproportionate weight to lexical cues when renaming preserves program structure. We introduce Face/Off, a semantics-preserving identifier-renaming framework, and evaluate progressive naming conditions across multiple models and code-comprehension tasks. Within this framework, lexical overemphasis is pervasive across the evaluated models and primary tasks: performance generally decreases as identifier information is removed or made misleading, and outputs are often directed toward the meanings suggested by misleading names. The pattern persists under representative prompt- and fine-tuning-based interventions, suggesting that lexical overemphasis is an entrenched problem. A type-inference control confirms a boundary: naming effects are smaller when the answer is locally recoverable without the target name. These results do not imply that identifiers are unhelpful; rather, they reveal a systematic vulnerability in how current LLMs balance lexical cues against program structure. Our findings motivate evaluations and modeling methods that preserve the benefits of natural code regularities while keeping conclusions grounded in accurate, formalized code semantics.
\end{abstract}



\maketitle
\begin{center}
\small\textit{Preprint. Manuscript under review.}
\end{center}
\section{Introduction}
Large language models (LLMs) are now widely used for source-code tasks~\cite{codebert,codet5,codex,xu2022systematic}, including code search~\cite{li2024rewriting}, code summarization~\cite{ahmed2024automatic}, clone detection~\cite{khajezade2024investigating}, bug localization~\cite{hossain2024deep}, and code completion~\cite{li2024attribution}. Recent instruction-following models have further improved performance, particularly in generative settings, and have become an important class of methods for both code understanding and generation.

These advances are consistent with the naturalness hypothesis: human-written code is repetitive and statistically predictable, which makes language modeling useful for software artifacts~\cite{hindle2012naturalness,allamanis2018survey}. Source code also contains natural-language information, particularly in identifiers. We use \textbf{lexical information} to refer to the information conveyed by the words in those identifiers. Such information is often useful rather than spurious by default: descriptive names can support human comprehension~\cite{schankin2018identifiers}, and high-quality lexical information and consistent coding styles can also facilitate models' processing of well-written code~\cite{jiang2024surveylargelanguagemodels,DataQualityForCodesummary}.

\begin{figure}[ht]
  \centering
  \includegraphics[width=\linewidth]{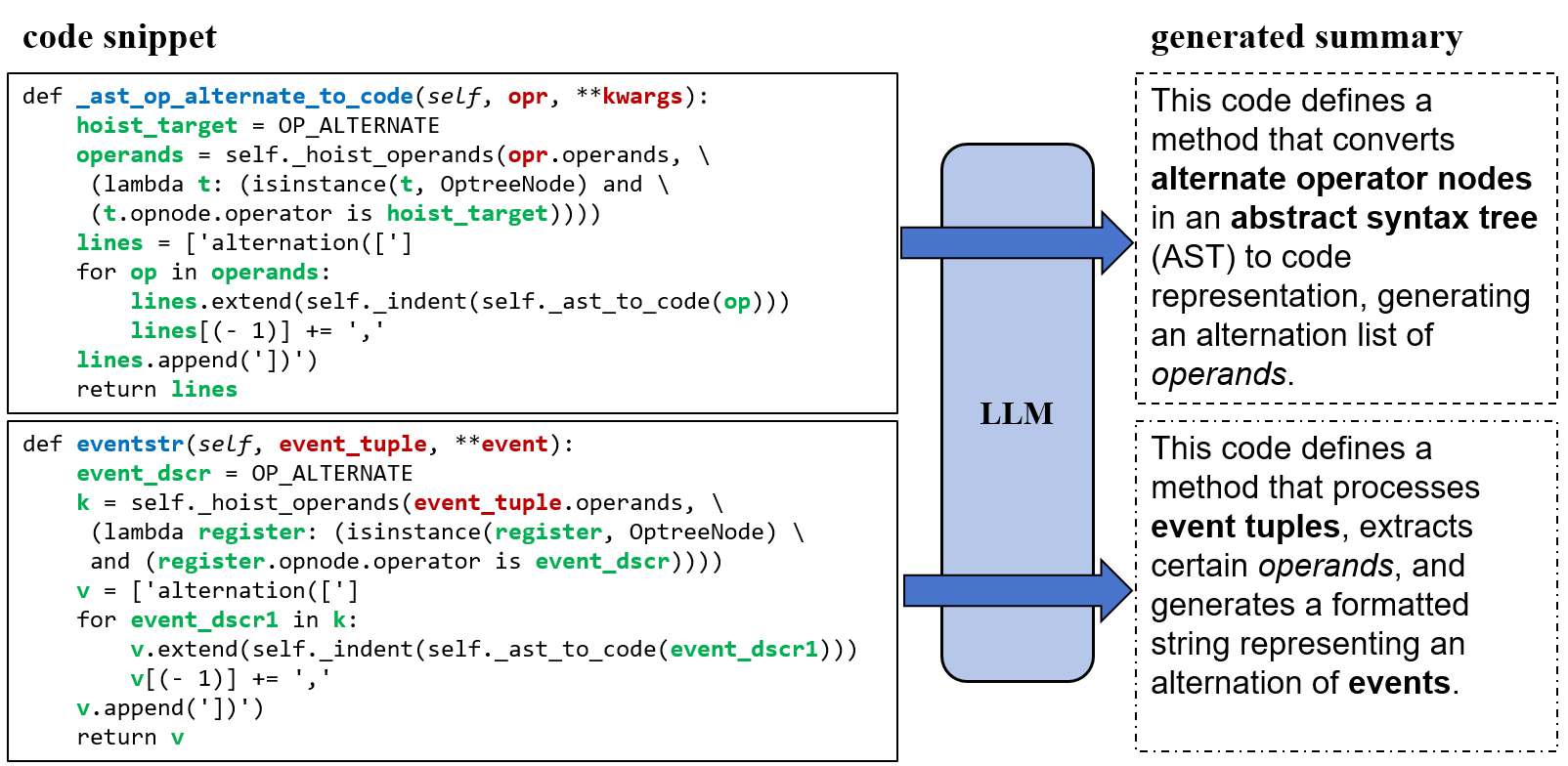}
  \caption{Code summaries generated by GPT-3.5-turbo for the same implementation under different naming conditions. Bold text marks output phrases associated with identifiers in the code, and italic text marks phrases associated with method names. In the lower example, the summary follows the misleading identifiers even though they conflict with the implementation.}
  \Description{Code summaries generated by GPT-3.5-turbo for the same implementation under different naming conditions. Bold and italic text mark phrases associated with identifiers in the code; in one example, the generated summary follows misleading names rather than the implementation.}
  \label{fig:case0}
\end{figure}

The formal semantics of a program, however, are not identical to the distributional regularities of the corpus in which it appears. For identifiers eligible for consistent alpha-renaming, lexical content can change without changing program behavior. Prior work has shown that code models can nevertheless be fragile under lexical perturbations~\cite{yan_towards_2021,yang_how_2024}. Figure~\ref{fig:case0}, for example, shows summaries that follow identifier meanings even when those meanings conflict with the implementation. We call this behavior \textit{lexical overemphasis} (or, metaphorically, \textit{lexical superstition}): disproportionate sensitivity to identifier-level cues when task-relevant program structure is held constant. Our central questions are therefore: Do the evaluated LLMs overemphasize lexical information during code comprehension, and, if so, to what extent and in what manner?

This question matters because benchmark performance may not fully characterize behavior when naming quality differs from the benchmark distribution. If a model assigns excessive weight to identifier cues, its reliability may decline on poorly named, anonymized, decompiled, or deliberately misleading code.

Because LLMs are trained on observational corpora, exploiting correlations between names and code behavior is expected and often beneficial. The reliability question is whether this dependence remains calibrated when the cues are missing or conflict with the implementation. This distinction matters because strong benchmark performance may be interpreted as evidence of code understanding or reasoning~\cite{gu2024cruxeval,li2025codei,ni2024next}, while models are increasingly considered for consequential tasks such as code review and vulnerability detection. Controlled counterexamples can reveal failure modes that ordinary test distributions may not expose.

Dependence is therefore not a binary attribute. Our purpose is not to reject software naturalness or to require models to ignore identifiers. Instead, we systematically characterize how model behavior changes as lexical cues move from informative to absent and then misleading, while data flow, control flow, and algorithmic structure remain unchanged.

We investigate this question with \textbf{Face/Off}, an evaluation framework based on semantics-preserving identifier transformations. We characterize how pervasive lexical sensitivity is across the evaluated models and tasks, test its persistence under representative interventions, and discuss possible contributing mechanisms more tentatively. To distinguish the loss of a task-relevant lexical cue from reliance on a cue that is unnecessary for the answer, we additionally conduct a targeted type-inference control on samples whose types can be determined from local program evidence.

\section{Related Work}

\subsection{LLMs for Code}

Learning-based methods for code have evolved from statistical and recurrent models to Transformer-based pre-trained models and instruction-following LLMs. These models are used in tasks such as code generation, completion, and vulnerability detection.

Early Transformer-based approaches adapted language models to code through additional pre-training, as in CodeBERT~\cite{codebert} and Codex~\cite{codex}. Other work incorporates program structure, including graph, control-flow, and data-flow information~\cite{cdlh,GraphCodeBERT,mayifancontrolflow}. More recent instruction-following model families, including GPT~\cite{tang2023chatgpt,gpt4}, DeepSeek~\cite{guo2024deepseek,zhu2024deepseek}, StarCoder~\cite{li2023starcoder}, and Qwen~\cite{qwen2,yang2024qwen2}, support a broad range of code tasks in zero- and few-shot settings.

Despite this progress, code models retain limitations in generation accuracy, robustness~\cite{Adversarial_robustness_for_code}, security and data contamination~\cite{Security_and_Privacy_llm,magar2022data}, interpretability, and data bias~\cite{jiang2024surveylargelanguagemodels}.

\subsection{Code Naturalness, Shortcut Learning, and Program Comprehension}

The naturalness hypothesis holds that human-written software is sufficiently repetitive and predictable to be modeled statistically~\cite{hindle2012naturalness}; this observation underpins much work on machine learning for code~\cite{allamanis2018survey}. It does not, however, imply that every distributionally predictive feature is necessary for a task's semantics. For a conditional model $p(y\mid x)$, identifier words may improve prediction because they correlate with behavior in ordinary corpora, even when a consistent renaming leaves the relevant program behavior unchanged. Our study does not dispute the usefulness of these correlations. It examines their reliability when lexical cues and program structure are experimentally placed in conflict.

This distinction parallels research on shortcut learning. Across machine learning, models can perform well on standard test distributions by using features that cease to be reliable under controlled or shifted conditions~\cite{geirhos2020shortcut}. In natural-language inference, for example, controlled counterexamples have been used to distinguish correct predictions based on intended relations from predictions based on fallible lexical and syntactic heuristics~\cite{mccoy2019wrong}. Face/Off applies the same diagnostic principle to code: semantics-preserving renaming holds the program structure fixed while varying one family of predictive cues. This connection motivates our experimental logic, but we limit our claims to the evaluated tasks, models, and transformations.

Identifier information is also important in human program comprehension. Descriptive identifiers can accelerate semantic-defect detection, with effects that depend on the comprehension task and developer experience~\cite{schankin2018identifiers}. Program-comprehension research likewise shows that programmers use recognizable code ``beacons,'' that misleading beacons can induce false initial interpretations~\cite{wiedenbeck1991initial}, and that procedural relations such as control flow contribute to experts' mental representations~\cite{pennington1987stimulus}. Field observations further show that professional comprehension extends beyond source text to browsers, documentation, and development tools~\cite{xia2018measuring}. Thus, neither humans nor models should be expected to ignore names. The relevant question is how conclusions are revised when a useful cue conflicts with other available evidence.

\subsection{Code Perturbation}

Code perturbation modifies selected properties of a program to observe a model's response, often while preserving functionality. Perturbation methods can be organized by model access (white-box or black-box) and by the transformed program feature, such as identifiers or coding style.

In code intelligence, code perturbation is widely used in attack-and-defense and robustness studies~\cite{du_extensive_2023}. Because white-box methods require model access that is not always available, many studies use black-box approaches. Notable identifier-based attacks include MHM~\cite{zhang_generating_2020}, ACCENT~\cite{zhou2022adversarial}, WIR-Random~\cite{zeng2022extensive}, and ALERT~\cite{yang_natural_2022}. Style-transfer methods insert dead code~\cite{na2023dip} or replace loop structures~\cite{li_ropgen_2022}.

Beyond adversarial contexts, code perturbation serves as a tool for probing model behaviors~\cite{dinh_large_2023, hooda_large_2024}. Lexical perturbation has recently gained attention: \citeauthor{yang_how_2024} examine its role in adversarial settings~\cite{yan_towards_2021, yang_how_2024, fang2024large}, \citeauthor{gao_two_2023} apply counterfactual analysis~\cite{gao_two_2023}, and \citeauthor{hu_how_2024} analyze naming quality~\cite{hu_how_2024}.

~

Although prior works have recognized the importance of lexical information for code comprehension, many study identifier perturbations in attack-and-defense or robustness settings. Our work is closely connected to this literature: we share semantics-preserving transformations, while using representative mitigation strategies in RQ3 as diagnostic interventions rather than as candidates for a state-of-the-art defense. Their role is to test whether the lexical influence characterized in RQ2 can be readily suppressed.

In RQ1 and RQ2, identifier renaming is a controlled probe for characterizing how the quality and content of lexical information affect model interpretation. Rather than optimizing an attack or a defense, we compare a progressive spectrum of naming conditions and analyze both the magnitude and direction of changes in model responses. These analyses examine whether model behavior is anchored more strongly in identifier-level cues than in the unchanged program structure.

The remaining performance gaps show that lexical overemphasis persists across the tested models and methods, without implying that every possible intervention must fail.

Accordingly, our contribution is complementary to attack-and-defense research. We borrow relevant perturbation and mitigation techniques but employ them in a progressive, diagnostic study of lexical reliance.

\section{Experimental Setup}

\subsection{Research Questions}

We organize the study around three research questions. RQ1 measures sensitivity to naming quality across the evaluated models and tasks; RQ2 examines whether misleading names directionally affect model outputs; and RQ3 uses mitigation attempts to probe the persistence of the observed sensitivity.

~

\noindent\textbf{RQ1:} \textbf{To what extent does identifier naming quality influence LLM performance on the evaluated code-comprehension tasks?}

RQ1 compares performance under progressive naming conditions while holding the eligible program structure and behavior fixed. It tests how broadly the pattern recurs across the models, architectures, and tasks included in this study.

~

\noindent\textbf{RQ2:} \textbf{In what manner does identifier information influence LLM code comprehension?}

Performance degradation alone does not show how names affect a model's interpretation. RQ2 therefore tests whether outputs move in the direction suggested by misleading identifiers, rather than merely changing under a distribution shift. Figure~\ref{fig:case0} provides an initial example; the experiments examine whether related directional patterns appear in the evaluated tasks.

~

\noindent\textbf{RQ3:} \textbf{What do mitigation attempts reveal about the persistence of lexical overemphasis in current LLMs?}

RQ3 extends RQ2 by testing whether representative prompt- and fine-tuning-based interventions readily suppress the observed lexical influence. We monitor both overall performance and the gap between naming conditions; a persistent gap indicates that the behavior is not readily mitigated by the evaluated interventions. The scope of this inference is discussed in Threats to Validity.

\subsection{Terminologies}\label{sect:term}

This section defines the terms used in our experimental design.

\subsubsection{Semantics of Code}\label{sect:semantics}

In this study, the relevant preservation criterion is contextual operational behavior: an eligible transformation leaves control flow, data flow, operations, literals, and interactions with the surrounding context unchanged. We use \textbf{functionality} more narrowly for the observable input--output mapping, including relevant effects on external state; two implementations may therefore provide the same functionality while differing internally. For the cases considered here, controlled alpha-renaming preserves both. We exclude boundary cases in which names can affect dynamic resolution or external interaction, and verify that no other program construct changes by checking AST identity modulo the renamed identifier tokens.

\subsubsection{Lexical Information of Code}\label{sect:lexical} 

The \textbf{lexical information} in this study is the additional information conveyed by the natural-language words in identifiers. Identifiers have both a nominal role in the program and a human-facing descriptive role. For identifiers that satisfy the eligibility rules of Face/Off, consistently renaming a binding and all of its references preserves the program's operational semantics, although it can change readability and the expectations of a human or model reader.

We therefore use \textit{identifier naming information} as an alternative term for lexical information. Calling this information semantics-preserving under renaming does not mean that names are useless; it means that, for the transformations studied here, their descriptive content is not required to preserve the executed behavior.

\subsubsection{"Semantics" of Non-Self-Contained Code Snippets}\label{sect:non-closed}

We call a snippet \textbf{self-contained} when the bindings needed to interpret the transformed code are defined within the supplied snippet. A non-self-contained snippet may depend on external functions, attributes, or variables whose behavior cannot be determined from the snippet alone. Such examples are common in datasets including CodeSearchNet~\cite{codesearchnet}. Models may use names and surrounding context as evidence about these unresolved dependencies, but the supplied snippet does not by itself establish their behavior. We therefore restrict the standard Face/Off transformation to eligible locally bound identifiers and examine external identifiers separately in Section~\ref{sect:extend}.

\subsection{Method: Face/Off}\label{sect:method}

Based on these considerations, we design an evaluation framework called \textbf{Face/Off} to examine how model behavior changes under controlled variation in identifier information. For identifiers that meet the eligibility conditions, consistent renaming preserves operational behavior while changing the lexical cues available to the model.

\begin{figure}[htbp]
\centering
\includegraphics[width=\linewidth]{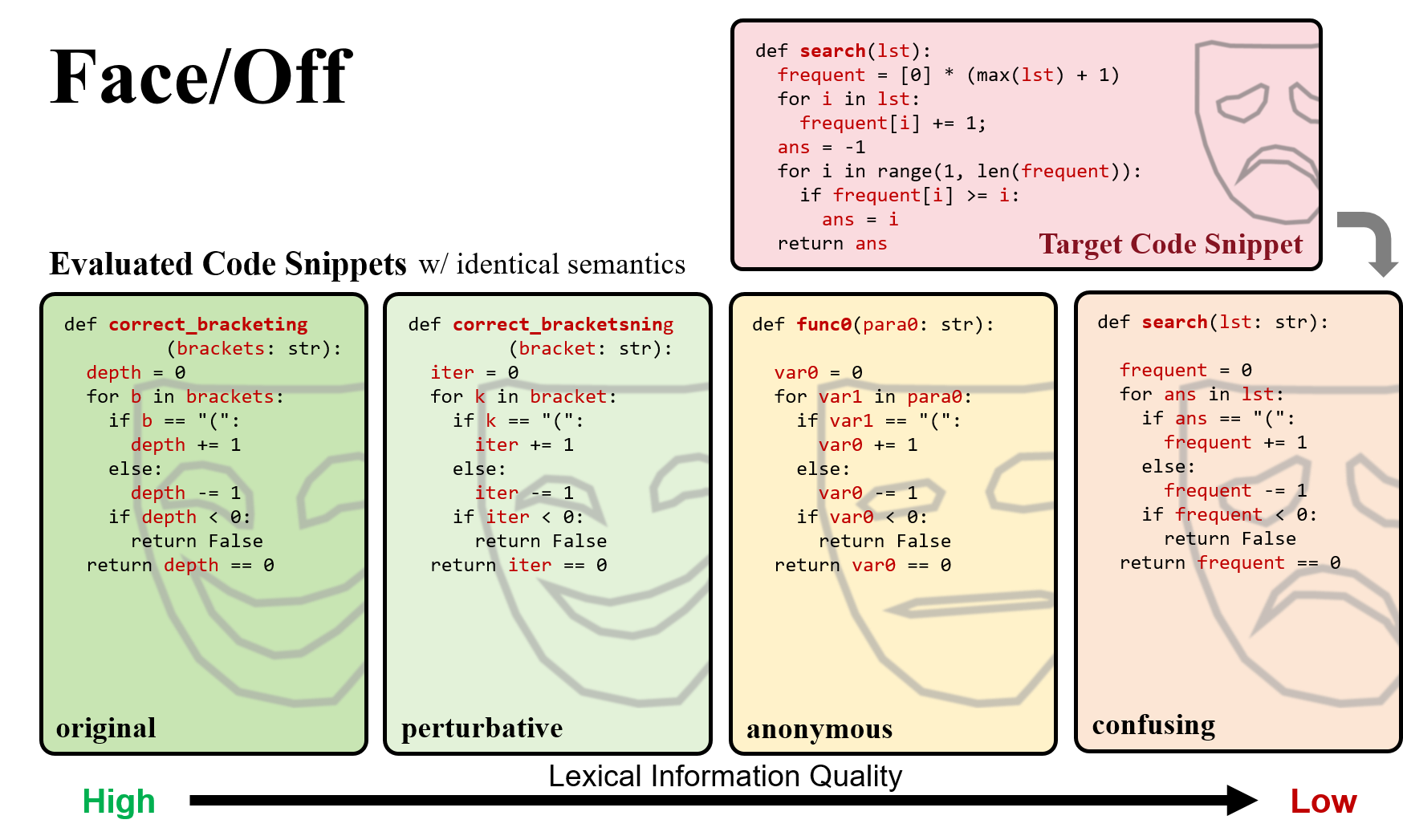}
\caption{The Face/Off evaluation framework, illustrated with a code snippet under different naming schemes. The Confusing condition uses identifiers from an unrelated donor snippet. For eligible renamings, the lexical ``face'' changes while operational behavior is preserved.\label{fig:case1}}
\Description{The schematic diagram of the Face/Off evaluation framework.}
\end{figure}

\textbf{Perturbative naming}. This condition replaces identifiers with semantically similar alternatives and retains naturalness as a selection criterion. We sample replacements using CodeBERT~\cite{codebert} representations rather than optimizing them against the evaluated model through M--H sampling~\cite{zhang_generating_2020} or evolutionary search~\cite{yang_natural_2022}. It provides a reference for relatively mild lexical variation.

\textbf{Anonymous naming}. This condition replaces eligible identifiers with systematic placeholders such as \texttt{func0}, \texttt{func1}, \texttt{var0}, and \texttt{var1}. The placeholders remove most descriptive content without intentionally pointing toward an unrelated function, providing a reference between informative and misleading names.

\textbf{Confusing naming}, or targeted naming. This condition tests whether identifiers drawn from another snippet can direct model outputs toward the meanings associated with that snippet. We call the donor snippet the \textit{target code} and construct the mapping using identifier type and order of appearance.

\textbf{Shuffling naming}. This scheme is used only for the exploratory analysis in Section~\ref{sect:shuffling}; it is not part of the standard Face/Off evaluation. It permutes the identifier mapping while retaining the same identifier set.

~

As illustrated in Figure~\ref{fig:case1}, Face/Off presents the same eligible program under progressively less informative and then misleading naming conditions. Its lexical ``face'' changes while the controlled alpha-renaming preserves operational behavior.

We denote by $\{C_A, D_A\}$ the source code and its corresponding documentation. Let $C_A^B$ denote the result of confusing renaming of $C_A$ using $C_B$ as the target code. $\mathcal{M}(C)$ denotes the model's output on $C$.

Accompanying text is handled by task rather than by renaming every identifier-like word. For the CodeSearchNet-based tasks, AST-based mappings transform the code; when a naming condition changes the subject-function name, an exact occurrence of that name in the corresponding description or reference follows the same mapping. Parameter names and other prose remain unchanged because they may convey task-relevant information. In code--document matching, only the positive description is synchronized; random negative descriptions remain unchanged. Transformation failures were manually completed so that all conditions retain the same 600 samples, and the resulting descriptions were inspected.

For HumanEval, the natural-language problem description is excluded from the input. The model receives only the transformed function signature and a truncated body prefix; the same mapping covers definitions and references in the prefix and held-out completion, and the evaluation entry point follows the transformed function name. Unit tests are withheld and used only for functional evaluation.

~

Confusing renaming introduces identifiers from a randomly selected donor snippet $C_B$ while retaining the eligible implementation of $C_A$; donors come from a different class when labels are available, and identifiers are mapped by type and order of appearance. We do not impose a subjective semantic-dissimilarity criterion, so incidental similarities are part of the sampling procedure and may weaken or strengthen individual contrasts. Section~\ref{sect:guiding} tests whether outputs move toward the meanings associated with the sampled donors.

~

In the standard Face/Off setting, external identifiers remain unchanged because their bindings and behavior may not be available in the snippet. Section~\ref{sect:extend} examines them separately with the extended construction and an insertion control.

\subsection{Tasks}

\paragraph{Code-Document Matching}

Code-Document Matching asks whether a code snippet and a brief description correspond. Given a pair $(C_i,D_j)$, the model outputs a binary decision. We use the original pair $(C_A,D_A)$ as a positive sample and a randomly selected irrelevant description $D_B$ to construct the negative pair $(C_A,D_B)$; text handling follows Section~\ref{sect:method}. As shown in Table~\ref{tab:results_match}, the evaluated models obtain a $100\%$ true negative rate (TNR) for these random pairs under the Original condition, indicating that the sampled negatives are readily distinguishable in that setting.

We report classification accuracy, along with true positive rate (TPR) and true negative rate (TNR). The CodeSearchNet dataset~\cite{codesearchnet} is used here, as well as for Code Search and Code Summarization. To prevent interference from other information, we modify comments by removing structured parameter-list sections, usage examples, URLs, and other content likely to provide direct answer cues.

\paragraph{Code Search}

Code Search evaluates a model's ability to retrieve code from a natural-language query. Given $D=\{(C_i,D_i)\}$, we encode code and text separately and rank each $C_j$ for query $D_i$ by the similarity between $\mathcal{M}(C_j)$ and $\mathcal{M}(D_i)$.

Mean Reciprocal Rank (MRR) is used as the evaluation metric~\cite{CodeXGLUE,Mrruse}, which measures the average of the reciprocal ranks of results for a set of queries. 

\paragraph{Code Summarization}

Code summarization requires the model to generate a one-sentence natural-language summary for a given code snippet $C_i$. Unlike code--document matching, it evaluates an open-ended description rather than a binary decision.

Following prior work~\cite{CodeXGLUE,summarymetricuse}, we evaluate the output $\mathcal{M}(C_i)$ against the reference description $D_i$ using BLEU~\cite{papineni_bleu_2002} (with 4-gram precision), ROUGE-L~\cite{rouge_2004} (with F1 reported as the primary metric), and METEOR~\cite{meteor_2005}. Here, $D_i$ is an evaluation reference rather than part of the model input, and its treatment follows Section~\ref{sect:method}.

\paragraph{Code Completion}

In Code Completion, the model receives a transformed function signature followed by a truncated prefix of the function body and generates the remainder of the solution. Input transformation and functional evaluation follow Section~\ref{sect:method}.

Functional correctness is evaluated using Pass@1~\cite{guo2024deepseek,codellama,Language_models_arefewshot,gpt4}. We use the HumanEval dataset~\cite{codex}. Unlike the standard setting, we filter examples with sufficiently long solution code, randomly truncate the code, and append the first few lines of the canonical solution to the input so that both the supplied prefix and held-out completion contain at least one identifier.

\subsection{Evaluated Models}

We evaluate a diverse set of models, including both large instruction-following LLMs (via in-context learning) and smaller open-source models (via fine-tuning).

\paragraph{Larger instruction-following models (in-context learning):}

\begin{itemize}
\item \texttt{GPT-3.5-turbo-0125}~\cite{gpt3.5}: A decoder-only model fine-tuned with RLHF for instruction following. We access it via OpenAI's API.

\item \texttt{GPT-4o-mini}~\cite{gpt4}: A representative of the GPT-4 family, known for strong performance across software engineering tasks.

\item \texttt{LLaMA3:70B-Instruct}~\cite{touvron2023llama}: A multilingual LLM trained on 1.4 trillion tokens, with 4.5\% code from GitHub.

\item \texttt{DeepSeek-Coder-V2}~\cite{zhu2024deepseek}: An open-source MoE code model with 236B total parameters (21B active), further pre-trained on 6 trillion tokens.
\end{itemize}

\paragraph{Smaller open-source models (fine-tuning):}

\begin{itemize}
\item \texttt{CodeBERT}~\cite{codebert}: A Transformer model pre-trained with masked language modeling and replaced token detection for code understanding.

\item \texttt{GraphCodeBERT}~\cite{GraphCodeBERT}: Extends CodeBERT with data flow information to encode variable relationships.

\item \texttt{CodeT5}~\cite{codet5}: An encoder-decoder model pre-trained on large-scale source code for both understanding and generation tasks.

\item \texttt{DeepSeek-Coder-1.3B}~\cite{guo2024deepseek}: A base model (1.3B parameters) pre-trained from scratch on 2 trillion tokens with a 16K window for code generation and infilling.
\end{itemize}

\subsection{Prompting}

We evaluate instruction-following LLMs through prompting, following their standard mode of use and without task-specific parameter updates.

\begin{figure}
    \centering    \includegraphics[width=\linewidth]{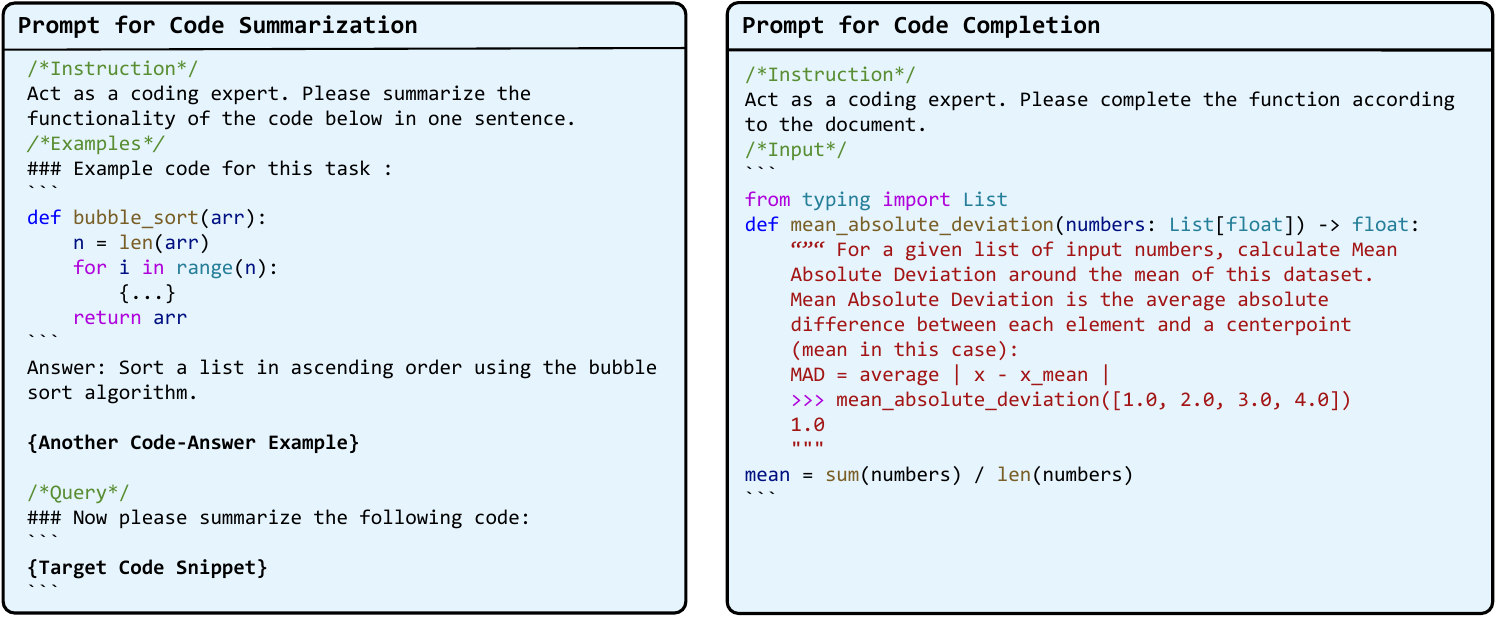}
    \caption{Two (abbreviated) examples of the prompt used in our experiments.}
    \label{fig:prompt}
    \Description{Two abbreviated examples of the prompt used in our experiments.}
\end{figure}

For code--document matching and code summarization, we use few-shot prompting as illustrated in Figure~\ref{fig:prompt}. Prompt examples are randomly sampled from the training set. Unless otherwise specified, each prompt contains two examples; for code--document matching, these comprise one positive and one negative example. Following prior work~\cite{nijkamp2022codegen}, we use the zero-shot setup shown in Figure~\ref{fig:prompt} for code completion.

\section{RQ1: Cross-Task Evaluation}
\label{Cross-Task Evaluation}

\subsection{Face/Off Main Results}

We now summarize the setting of cross-task Face/Off evaluation with Table~\ref{tab:overview}, following the ideas and criteria mentioned above.

\begin{table}[htbp]
\centering
\caption{This table illustrates the correspondence between tasks, models and datasets used in this section. The upper section lists large language models evaluated using in-context learning, while the lower section details open-source models evaluated through fine-tuning.}\label{tab:overview}
\footnotesize
\begin{tabular}{cccccc}
\toprule

\multirow{2}{*}{\textbf{Evaluated Models}}& \textbf{Task} & Code-Doc Matching & Code Search & Code Summarization & Code Completion \\
\cline{2-6}
& \textbf{Dataset} & CodeSearchNet  & CodeSearchNet& CodeSearchNet& HumanEval \\
\midrule
\multicolumn{2}{l}{\;\;GPT-3.5-turbo}&\checkmark& &\checkmark&\checkmark\\
\multicolumn{2}{l}{\;\;LLaMA3:70B-Instruct}&\checkmark& &\checkmark&\checkmark\\
\multicolumn{2}{l}{\;\;DeepSeek-Coder-V2}&\checkmark& &\checkmark&\checkmark\\
\multicolumn{2}{l}{\;\;GPT-4o-mini}&\checkmark& &\checkmark&\checkmark\\
\multicolumn{2}{l}{\;\;Text-embedding-3}& &\checkmark & & \\
\midrule
\multicolumn{2}{l}{\;\;CodeBERT}& &\checkmark&\checkmark&\\
\multicolumn{2}{l}{\;\;GraphCodeBERT}& &\checkmark&\checkmark&\\
\multicolumn{2}{l}{\;\;CodeT5}& & &\checkmark&\\
\multicolumn{2}{l}{\;\;DeepSeek-Coder-1.3B}& & &\checkmark&\\
\bottomrule
\end{tabular}
\end{table}

For each task, the naming conditions are derived from the same underlying samples. The controlled transformation is designed to vary eligible identifier information while retaining the other program content, thereby isolating this factor within the limits discussed in Section~\ref{sect:method} and Threats to Validity.

\subsubsection{Code-Document Matching}

Table~\ref{tab:results_match} reports code-document matching accuracy. For each evaluated model, performance decreases from Original to Anonymous naming and decreases further under Confusing naming; Perturbative naming produces a smaller change.

\begin{table}[htbp]
\centering
\caption{Performance (shown in accuracy \%) of various LLMs under different Face/Off schemes. Subscripts indicate corresponding true positive and true negative rates (TPR$\vert$TNR).}\label{tab:results_match}
\footnotesize
\begin{tabular}{lcccc}
\toprule
\multirow{2}{*}{\textbf{Evaluated Model}}& \multicolumn{4}{c}{\textbf{Setting}}\\
\cline{2-5}
 & \textbf{Original} &\textbf{Perturbative}  &\textbf{Anonymous}   &\textbf{Confusing}  \\
\midrule
\textbf{GPT-3.5-turbo} 
 & 78.27$_{56.54\vert100.0}$ & 71.46$_{42.93\vert100.0}$ & 67.61$_{35.23\vert100.0}$ & 57.16$_{14.99\vert99.32}$ \\

\textbf{LLaMA3:70B-Instruct}
 & 93.96$_{87.92\vert100.0}$ & 91.77$_{83.71\vert99.83}$ & 87.56$_{75.13\vert100.0}$ & 70.61$_{59.28\vert81.94}$ \\

\textbf{DeepSeek-Coder-V2} 
 & 90.52$_{81.04\vert100.0}$ & 86.58$_{73.16\vert100.0}$ & 85.26$_{70.70\vert99.83}$ & 73.08$_{47.36\vert98.81}$ \\

\textbf{GPT-4o-mini} 
 & 88.26$_{76.51\vert100.0}$ & 84.65$_{69.30\vert100.0}$ & 79.73$_{59.45\vert100.0}$ & 67.89$_{39.18\vert96.59}$ \\
\bottomrule
\end{tabular}
\end{table}

Examining TPR and TNR separately, TNR remains near $100\%$ in most conditions, whereas TPR accounts for most of the accuracy decline. The lower TNR under Confusing naming for some models also shows that donor identifiers can make a mismatched description appear more plausible. Section~\ref{sect:guiding} examines this directional effect more directly.

\subsubsection{Code Search}

Table~\ref{tab:main_result_code_search} presents the MRR results for code search. Among the evaluated models, CodeBERT degrades the most, while GraphCodeBERT---using the same architecture but with additional data flow-based attention---performs similarly to the commercial-grade encoder Text-embedding-3~\cite{openai_textembedding3}.

\begin{table}[htbp]
\centering
\small
\caption{Mean Reciprocal Rank (MRR) results for code search and performance degradation ratios for two fine-tuned models and the encoding API under different Face/Off naming schemes.}\label{tab:main_result_code_search}
\begin{tabular}{lcccc}
\toprule
\multirow{2}{*}{\textbf{Evaluated Model}}& \multicolumn{4}{c}{\textbf{Setting}}\\
\cline{2-5}
 & \textbf{Original} &\textbf{Perturbative}  &\textbf{Anonymous}   &\textbf{Confusing}  \\
\midrule
\textbf{CodeBERT} 
 & 93.05 & 84.19 (9.521\%\textcolor{red}{\ensuremath{\downarrow}}) & 72.77 (21.79\%\textcolor{red}{\ensuremath{\downarrow}}) & 37.06 (60.17\%\textcolor{red}{\ensuremath{\downarrow}}) \\

\textbf{GraphCodeBERT} 
 & 93.98 & 86.78 (7.661\%\textcolor{red}{\ensuremath{\downarrow}}) & 79.81 (15.07\%\textcolor{red}{\ensuremath{\downarrow}}) & 45.82 (51.24\%\textcolor{red}{\ensuremath{\downarrow}}) \\

\textbf{Text-embedding-3} 
 & 91.48 & 78.13 (14.59\%\textcolor{red}{\ensuremath{\downarrow}}) & 72.15 (21.13\%\textcolor{red}{\ensuremath{\downarrow}}) & 47.21 (48.39\%\textcolor{red}{\ensuremath{\downarrow}}) \\
\bottomrule
\end{tabular}
\end{table}

All three renamed variants reduce MRR, with the largest decrease under Confusing naming. Because retrieval is based on encoder similarity, these results are consistent with lexical changes altering the relative positions of code and queries in the representation space. MRR alone, however, does not identify the direction or magnitude of the underlying embedding shifts.

\subsubsection{Code Summarization}

Table~\ref{tab:main_result_code_summary} reports average code-summarization scores. Instruction-following models are evaluated through few-shot prompting, whereas the smaller open-source models are fine-tuned. Every evaluated model scores below its Original condition after renaming, and Confusing naming yields the lowest score in each case.

\begin{table}[htbp]
\centering
\small
\caption{ROUGE-L F1 scores for code summarization of various LLMs under Face/Off framework. The upper section shows instruction-following models evaluated via in-context learning, and the lower section presents fine-tuned open-source models.}\label{tab:main_result_code_summary}
\begin{tabular}{lccccc}
\toprule
\multirow{2}{*}{\textbf{Evaluated Model}}& \multicolumn{4}{c}{\textbf{Setting}}\\
\cline{2-5}
 & \textbf{Original} &\textbf{Perturbative}  &\textbf{Anonymous}   &\textbf{Confusing}  \\
\midrule
GPT-3.5-turbo& 18.75 &  14.08 & 14.97  & 12.76 \\
LLaMA3:70B-Instruct& 17.82 & 15.48 & 15.28   & 13.71 \\
DeepSeek-Coder-V2 & 19.92&17.34& 15.94 & 14.55 \\
GPT-4o-mini & 17.81 &15.33& 14.44   & 13.73 \\
\midrule
DeepSeek-Coder-1.3B&22.22&15.88&13.63&12.31\\
 CodeBERT &19.50&13.24&13.75&9.84\\
 CodeT5 &22.64&16.98&16.75&11.69\\
 GraphCodeBERT &20.06&15.87&12.98&9.66\\
\bottomrule
\end{tabular}

\end{table}

The same ordering appears for both instruction-following and fine-tuned models, although the absolute scores and the size of the decreases differ.

\subsubsection{Code Completion}\label{sect:complete}

Table~\ref{tab:main_result_code_complete} presents code completion performance of four LLMs across all Face/Off settings. 

\begin{table}[htbp]
\footnotesize
\centering
\caption{Pass@1 results for Code Completion and performance degradation ratios for evaluated LLMs under different Face/Off naming schemes.}\label{tab:main_result_code_complete}
\begin{tabular}{lcccc}
\toprule
\multirow{2}{*}{\textbf{Evaluated Model}}& \multicolumn{4}{c}{\textbf{Setting}}\\
\cline{2-5}
 & \textbf{Original} &\textbf{Perturbative}  &\textbf{Anonymous}   &\textbf{Confusing}  \\
\midrule
\textbf{GPT-3.5-turbo} 
 & 66.93  & 61.42(8.232\%\textcolor{red}{\ensuremath{\downarrow}}) & 65.35(2.360\%\textcolor{red}{\ensuremath{\downarrow}}) & 55.51(17.06\%\textcolor{red}{\ensuremath{\downarrow}})\\

\textbf{LLaMA3:70B-Instruct}
 & 62.20  & 52.28(15.94\%\textcolor{red}{\ensuremath{\downarrow}}) & 58.26(6.334\%\textcolor{red}{\ensuremath{\downarrow}}) & 40.94(34.18\%\textcolor{red}{\ensuremath{\downarrow}}) \\

\textbf{DeepSeek-Coder-V2} 
 & 86.22 & 83.07(3.653\%\textcolor{red}{\ensuremath{\downarrow}}) & 89.37(3.653\%\textcolor{green}{\ensuremath{\uparrow}}) & 57.48(33.33\%\textcolor{red}{\ensuremath{\downarrow}})\\

\textbf{GPT-4o-mini} 
 & 88.19 & 76.38(13.39\%\textcolor{red}{\ensuremath{\downarrow}}) & 83.85(4.921\%\textcolor{red}{\ensuremath{\downarrow}})  & 73.22(16.97\%\textcolor{red}{\ensuremath{\downarrow}}) \\
\bottomrule
\end{tabular}
\end{table}

Code completion contains the clearest exceptions to the ordering observed in the other tasks. Anonymous naming sometimes performs better than Perturbative naming, and DeepSeek-Coder-V2 improves slightly over its Original score under Anonymous naming.

We attribute this in part to the nature of the task. The identifiers being renamed constitute a relatively small portion of the supplied signature and partial function body, while literals, operations, and control/data-flow relations remain available. Because the HumanEval descriptions are not included in the model input, the reversal cannot be attributed to reliance on the natural-language problem statement. Anonymous naming supplies consistent neutral placeholders, whereas perturbative naming can introduce alternative lexical associations; we treat this explanation as a hypothesis rather than a separately identified causal result.

Confusing naming nevertheless produces the lowest Pass@1 for each evaluated model, with decreases ranging from 16.97\% to 34.18\% relative to Original.

~

Across the four tasks, removing or contradicting identifier information generally reduces performance, with task-specific exceptions in code completion. Section~\ref{sect:sum} summarizes these patterns and their scope.

\paragraph{Comments on Prompt Setting}\label{sect:prompt}

Prompt design can affect instruction-following models. As a sensitivity analysis, we evaluate GPT-3.5-turbo on code-document matching with zero-shot, one-shot, and two-shot configurations using different example combinations.

\begin{table}[htbp]
\centering
\small
\caption{\label{tab:detailed match task}Effect of prompt settings on GPT-3.5-turbo in code--document matching. We report average accuracy, with subscripts indicating the corresponding true-positive and true-negative rates (TPR$\vert$TNR).}
\begin{tabular}{lccc}
\toprule
  \textbf{Prompt setting}  & \textbf{Original}&   \textbf{Anonymous}&\textbf{Confusing}\\
\midrule[1pt]

\textbf{Zero-Shot}&    69.21$_{38.42\vert100.0}$& 63.29$_{26.57\vert100.0}$ &56.31$_{13.46\vert99.15}$\\
\textbf{One-Shot(+)}&  77.35$_{54.87\vert99.83}$& 70.70$_{41.57\vert99.83}$ &57.16$_{16.52\vert97.79}$\\
\textbf{Two-Shot(+/+)}&77.18$_{54.36\vert100.0}$& 67.98$_{35.95\vert100.0}$ &56.73$_{14.48\vert98.98}$\\
\textbf{Two-Shot(+/-)}&78.27$_{56.54\vert100.0}$& 71.47$_{42.93\vert100.0}$ &57.16$_{14.99\vert99.32}$\\

\bottomrule
\end{tabular}

\end{table}

As shown in Table~\ref{tab:detailed match task}, absolute accuracy varies across prompts, but all tested configurations retain the ordering Original $>$ Anonymous $>$ Confusing. This result shows that the pattern is not specific to the single prompt used in the main experiment; it does not establish invariance to prompt design more generally.

\subsection{Impact of External Identifiers}\label{sect:extend}

The standard experiments rename only eligible locally defined identifiers. External functions, attributes, and variables may also provide informative lexical cues, but their behavior is not defined within the snippet. We therefore examine them separately with an extended construction.

\paragraph{Method: Extended Face/Off}

As discussed in Section~\ref{sect:non-closed}, directly renaming unresolved external identifiers can change the referenced object. The extended construction uses an added binding block to preserve the original behavior. Because the inserted code changes the program representation, this construction is less strictly controlled than local alpha-renaming.

\begin{figure}[htbp]
\centering
\includegraphics[scale=0.55]{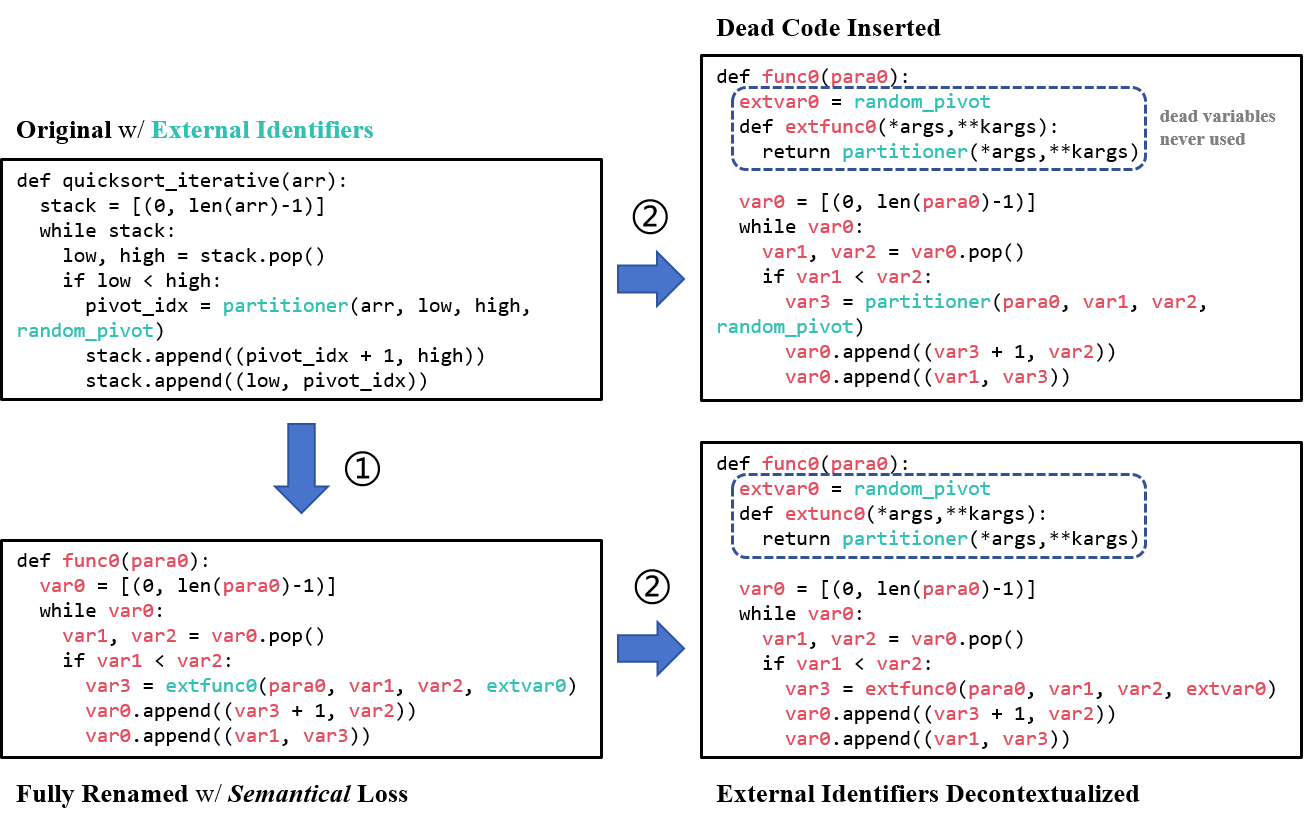}
\caption{An example of decontextualizing an external variable or function by wrapping it in a local definition. The corresponding dead-code insertion used for the control condition is also illustrated.
 }\label{fig:encapsulationcase}
 \Description{An example of our extended Face/Off together with the control group using dead code insertion.}
\end{figure}

Figure~\ref{fig:encapsulationcase} illustrates the extended construction. We rename an external identifier and insert a preceding assignment that binds the new local name to the original external object. This retains access to the same object while moving the original lexical cue out of its use sites.

The original external name therefore remains once in the binding block but is removed from its original use sites. This decontextualization tests whether relocating the cue changes model performance.

Because the added binding block changes the syntax, the extended method is not directly comparable to the local alpha-renaming intervention. We include a control that inserts a structurally similar dead-code block without renaming~\cite{na2023dip}. This control estimates part of the insertion effect, although it cannot remove every difference between the conditions.

\paragraph{Experiment Results}

We employ the extended Face/Off to evaluate various models on two tasks: code-document matching and code summarization. The results are shown in Table~\ref{tab:encapsulation}.

\begin{table}[htbp]
\centering
\small
\caption{Re-evaluation on two tasks, illustrating further performance degradation from decontextualizing external variables. Performance in the Original condition provides a reference for the model's response to dead-code insertion.}\label{tab:encapsulation}
\begin{tabular}{llcc}
\toprule
\textbf{Evaluated Model}&\textbf{Setting} & \textbf{Code-Doc Matching} &\textbf{Code Summarization}\\
\midrule
\multirow{5}{*}{\textbf{GPT-3.5-turbo}}
&\textbf{Original}&78.27&18.75\\
&\;\;w/ Dead Code&64.00&19.04\\
&\textbf{Anonymous}&71.46&14.97\\
&\;\;w/ Dead Code&56.15&13.29\\
&\;\;w/ Extended Face/Off&53.70&12.73\\

\midrule
\multirow{5}{*}{\textbf{LLaMA3:70B-Instruct}}
&\textbf{Original}&93.96&17.82\\
&\;\;w/ Dead Code&93.10&16.75\\
&\textbf{Anonymous}&87.56&15.28\\
&\;\;w/ Dead Code&88.11&14.47\\
&\;\;w/ Extended Face/Off&85.86&13.73\\

\midrule
\multirow{5}{*}{\textbf{DeepSeek-Coder-V2}}
&\textbf{Original}&90.52&19.92\\
&\;\;w/ Dead Code&89.69&17.08\\
&\textbf{Anonymous}&85.26&15.94\\
&\;\;w/ Dead Code&86.60&14.16\\
&\;\;w/ Extended Face/Off&83.67&13.60\\

\midrule
\multirow{5}{*}{\textbf{GPT-4o-mini}}
&\textbf{Original}&88.26&17.81\\
&\;\;w/ Dead Code&87.75&17.63\\
&\textbf{Anonymous}&79.73&14.44 \\
&\;\;w/ Dead Code&79.34&13.68\\
&\;\;w/ Extended Face/Off&75.51 &13.47\\

\bottomrule
\end{tabular}

\end{table}

After external identifiers are decontextualized, the table shows additional performance decreases for every evaluated model and task relative to the corresponding Anonymous-plus-dead-code condition. The size of this additional decrease varies across models and tasks.

These results are consistent with external identifier names contributing to performance when they appear at their use sites. Because the intervention relocates rather than eliminates the names and introduces an added binding block, it provides supporting evidence rather than a clean estimate of the effect of all external identifiers. We therefore do not generalize this result to every identifier or external dependency.

\subsection{Targeted Type-Inference Semantic Control}\label{sect:type-control}

The original four tasks can legitimately reward identifier information: descriptions, queries, and completions may be easier when names express intent. Their performance changes therefore do not alone distinguish disproportionate lexical reliance from the removal of information that the task ordinarily uses. We add a targeted control in which a selected variable's type is already determined by literals, operations, collection construction, comparisons, or local control and data flow. This supplementary boundary-condition experiment asks whether the larger effects in the original tasks persist when the answer is locally recoverable without the target name; it is not a fifth task intended for direct cross-task comparison.

\paragraph{Data and construction.}

We use the clean test split of ManyTypes4Py v0.7~\cite{mt4py2021}. A conservative automatic screen retains targets whose assignments are locally type-determining without repository-level or external-API type knowledge. We remove direct type cues from annotations, type comments, comments, and function docstrings while retaining the surrounding program structure and non-target lexical context. The resulting set contains 317 target variables from 217 projects and 217 source files.

Each target yields three paired conditions. \textit{Original} retains its name; \textit{Target-Anonymous} replaces the target binding and its references with a collision-free placeholder; and \textit{Target-Confusing} replaces the same binding and references with a real identifier associated with a conflicting top-level type. No other identifier, API, literal, or program relation is changed. The variants pass parsing, compilation, binding-occurrence, and alpha-equivalence checks.

\paragraph{Models and evaluation.}

The prompt identifies one target variable in the sanitized code and asks for one of 15 normalized type labels. We evaluate \texttt{GPT-3.5-turbo-0125}, \texttt{GPT-4o-mini}, and \texttt{Llama-3.1-70B-Instruct}~\cite{dubey2024llama3}.\footnote{\texttt{Llama-3.1-70B-Instruct} was used as the closest available successor to the \texttt{LLaMA3:70B-Instruct} endpoint used in the main experiments.} All conditions use the same prompt and temperature 0. We report target-level strict exact-match accuracy; paired confidence intervals are clustered by project, and McNemar tests are Holm-corrected within each model.

\begin{table}[htbp]
\centering
\caption{Target-level strict exact-match accuracy for the type-inference semantic control. Each condition contains the same 317 targets.}\label{tab:type-control}
\footnotesize
\begin{tabular}{lccc}
\toprule
\textbf{Model} & \textbf{Original} & \textbf{Target-Anonymous} & \textbf{Target-Confusing} \\
\midrule
GPT-3.5-turbo-0125 & 90.5\% & 88.3\% & 89.0\% \\
GPT-4o-mini & 82.6\% & 79.5\% & 84.9\% \\
Llama-3.1-70B-Instruct & 90.2\% & 90.2\% & 87.4\% \\
\bottomrule
\end{tabular}
\end{table}

\paragraph{Results.}

Table~\ref{tab:type-control} shows changes of $-3.2$ to $+2.2$ percentage points relative to Original, with different directions across models and conditions. None of the Anonymous--Original or Confusing--Original differences is significant after Holm correction. The only corrected significant contrast is GPT-4o-mini's Confusing accuracy relative to Anonymous ($+5.4$ points; project-clustered 95\% CI $[+1.2,+9.8]$; Holm-adjusted $p=.018$), opposite to a prediction that misleading names must consistently reduce accuracy. Errors and condition changes are concentrated in \texttt{dict} targets, whereas many primitive targets are at or near ceiling.

Within this conservatively screened subset, where local semantic evidence suffices for the answer, changing only the target name has a smaller and less consistent effect than in the original four tasks. Names can still modulate individual predictions, particularly for composite types, but they are not the principal required evidence in this conservatively screened subset. This result confirms a boundary on the main finding rather than showing that models never use names or that the original tasks should be name-independent.

\subsection{Section Summary}\label{sect:sum}

\begin{figure}[ht]
  \centering
  \includegraphics[width=\linewidth]{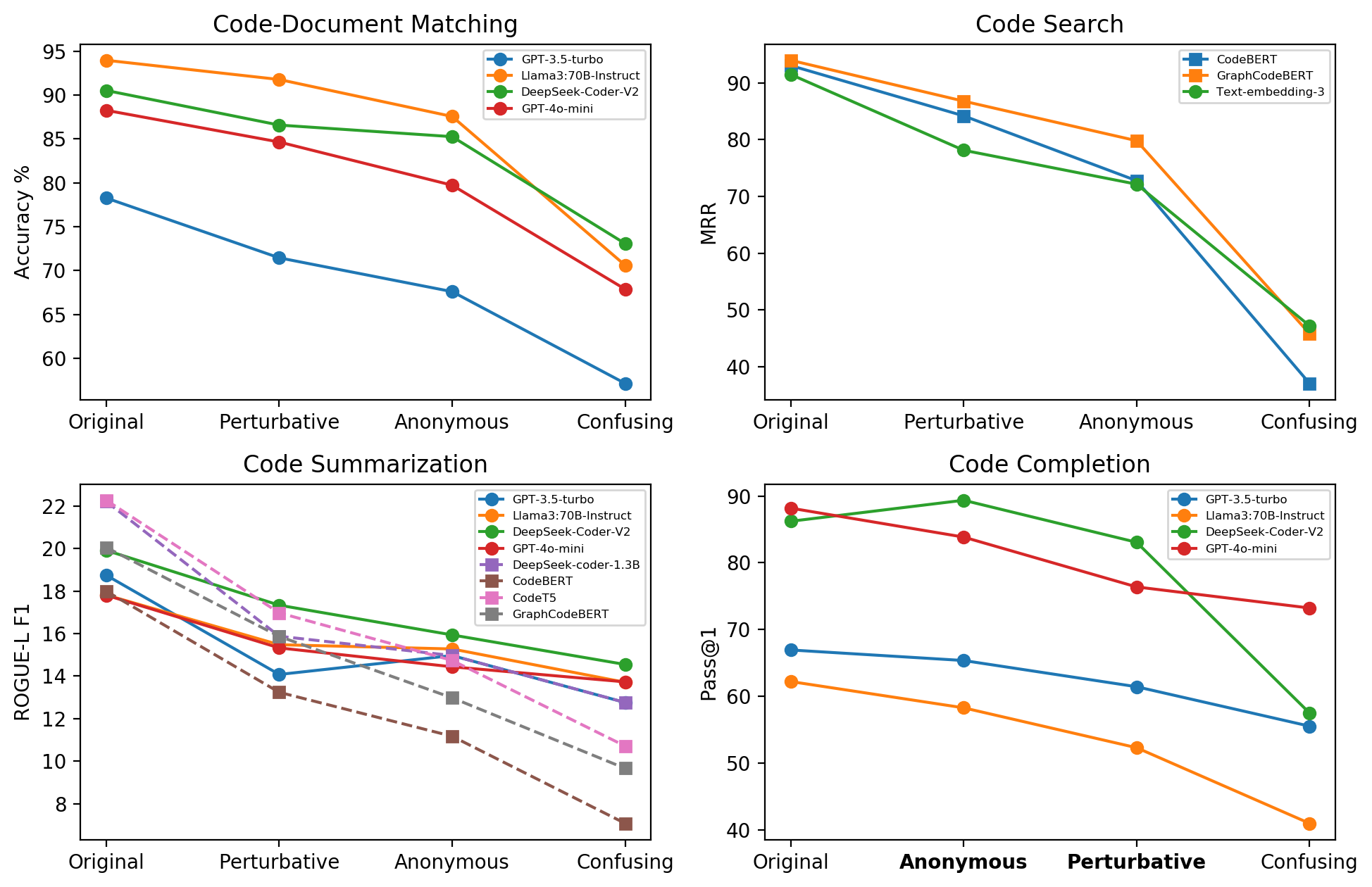}
  \caption{Line chart display of Face/Off evaluation results. In the figure, circular data points represent larger instruction-following models, while square data points represent smaller open-source models. Note that in the code completion chart, we have partially reversed the order, and the considerations are discussed in~\ref{sect:complete}.}
  \label{fig:overview}
  \Description{Line chart display of Face/Off evaluation results. The trend lines on each chart generally show a downward trend from left to right, with only a few exceptions.}
\end{figure}

Figure~\ref{fig:overview} summarizes the cross-task results. Perturbative naming generally produces the smallest decrease, Anonymous naming a larger decrease, and Confusing naming the largest. Code completion departs from this ordering for Perturbative versus Anonymous naming, but Confusing remains the lowest-performing condition for each evaluated model.

The extended experiment further suggests that the location and context of external identifier cues can matter, although its added binding block makes the evidence less controlled than the standard Face/Off transformations.

Overall, sensitivity to identifier quality is pervasive across the evaluated models and original four tasks. The targeted type-inference control confirms a boundary on its magnitude: in the conservatively screened subset, effects are smaller and directionally inconsistent when the answer is recoverable without the target name. RQ2 next asks whether the sensitivity observed in the original tasks is directional---that is, whether misleading names guide outputs toward the donor meaning---rather than treating performance changes alone as sufficient evidence of overemphasis.

\section{RQ2: Effect Analysis}

RQ1 shows performance sensitivity under controlled naming changes, but it does not by itself distinguish loss of a useful cue from active guidance by a misleading one. RQ2 addresses this distinction by testing whether outputs move toward the semantics suggested by donor identifiers.

\subsection{Guiding Effect of Lexical Information}\label{sect:guiding}

We focus on Confusing naming, where $C_A^B$ retains the eligible implementation of program $A$ but receives identifiers sampled from program $B$. RQ1 evaluates outputs against the correct target for $A$; here we additionally compare them with references associated with $B$. A shift toward $B$ provides directional evidence that the donor identifiers influence the output, although it does not by itself identify the model's internal mechanism.

\paragraph{Code-Document Matching}

Because negative samples in the original matching task have TNR near $100\%$, we reformulate the analysis as a two-document selection task. The model sees the correct and donor descriptions alongside the code and chooses between them. Table~\ref{tab:select task} shows lower accuracy under Confusing naming for every model; for GPT-3.5-turbo, accuracy falls from 92.84\% to 52.39\%, close to chance in this binary task. The smaller or absent changes under Anonymous naming support a directional effect from the donor names rather than a generic consequence of renaming alone.

\begin{table}[htbp]
\centering
\small
\caption{Performance metrics and degradation ratios of several LLMs across three naming schemes for code-document selection.}\label{tab:select task}
\begin{tabular}{lcccc}
\toprule
\multirow{2}{*}{\textbf{Setting}} & \multicolumn{4}{c}{\textbf{Evaluated Model}}  \\
\cline{2-5}
 & \textbf{GPT-3.5-turbo} & \textbf{LLaMA3:70B-Instruct} & \textbf{DeepSeek-Coder-V2} & \textbf{GPT-4o-mini} \\
\midrule 
Original & 92.84\% & 98.63\% & 98.98\% & 98.98\% \\
\cline{2-5}
Anonymous & 
81.35\%(12.37\%\textcolor{red}{\ensuremath{\downarrow}}) & 98.63\%(0.000\%\textcolor{gray}{-}) & 98.63\%(0.35\%\textcolor{red}{\ensuremath{\downarrow}}) & 98.98\%(0.000\%\textcolor{gray}{-}) \\
Confusing & 52.39\%(43.56\%\textcolor{red}{\ensuremath{\downarrow}}) & 73.27\%(25.71\%\textcolor{red}{\ensuremath{\downarrow}}) & 78.66\%(20.52\%\textcolor{red}{\ensuremath{\downarrow}}) & 76.11\%(23.10\%\textcolor{red}{\ensuremath{\downarrow}}) \\
\bottomrule
\end{tabular}
\end{table}

\paragraph{Code Summarization}

For summarization, let $d(\cdot,\cdot)$ denote a text-similarity metric such as BLEU. In addition to the standard score $d(\mathcal{M}(C_A^B),D_A)$ against the correct reference, we compute $d(\mathcal{M}(C_A^B),D_B)$ against the donor reference. We use $d(\mathcal{M}(C_A^B),D^\prime)$ for a randomly selected unrelated description $D^\prime$ as a background comparison. A donor-reference score above this random-pair level indicates that the change is directionally associated with the donor, rather than an arbitrary drift.

\begin{figure}[htbp]
\centering
\includegraphics[width=\linewidth]{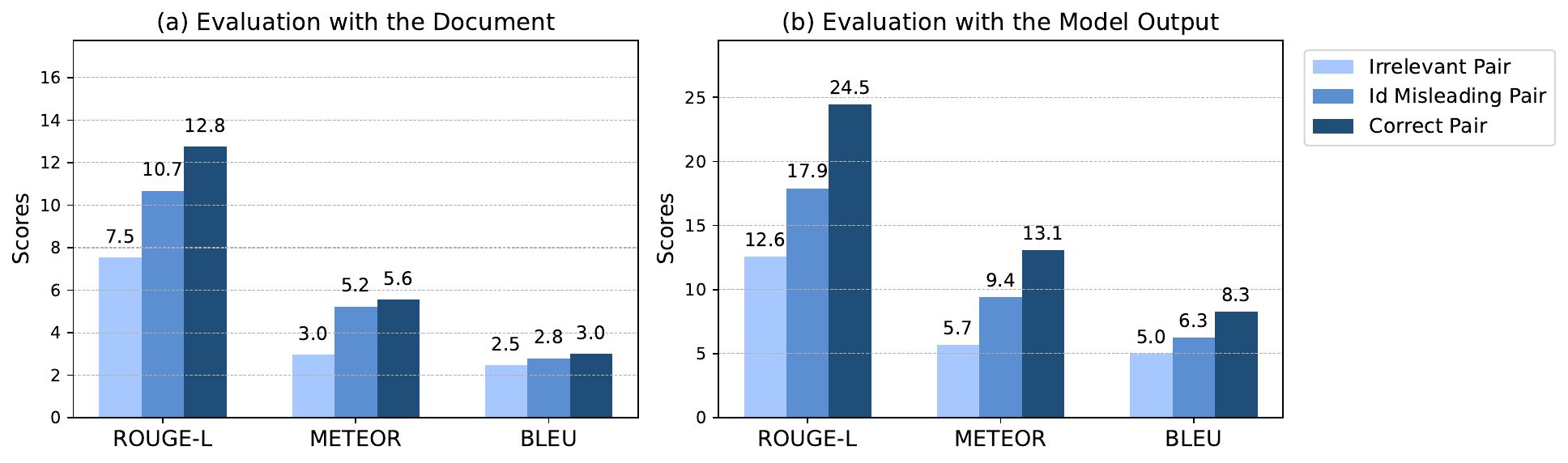}
\caption{Similarity comparisons for summaries generated under Confusing naming, using irrelevant, donor, and original pairs as references.} \label{fig:summary_attack_pair}
\Description{Similarity scores for summaries generated under Confusing naming. Donor-pair scores are higher than random-pair scores and lower than original-pair scores.}
\end{figure}

We repeat the comparison using model-generated summaries as references. The three quantities are the random-pair comparison, similarity to the Original output, and similarity to the donor output:

\[
\begin{split}
&d(\mathcal{M}(C),\mathcal{M}(C^\prime)),\\
&d(\mathcal{M}(C_A^B),\mathcal{M}(C_A)),\quad
d(\mathcal{M}(C_A^B),\mathcal{M}(C_B)).
\end{split}
\]

Figure~\ref{fig:summary_attack_pair} shows that donor-pair scores exceed the random-pair comparison for both human references and model-generated summaries, while remaining below the corresponding original-pair scores. This pattern is consistent with partial directional influence from the donor identifiers: the outputs do not fully switch to the donor functionality, but their lexical content moves toward it.

\paragraph{Code Completion}

For a qualitative code-completion analysis, we inspect samples that are correct under Original naming but incorrect after renaming and compare the generated continuations across conditions. This conditioning does not eliminate sampling variance, but it isolates examples in which the observed outcome changes and allows us to examine whether the error is related to the donor identifiers.

\begin{figure}[htbp]
\centering
\includegraphics[width=\linewidth]{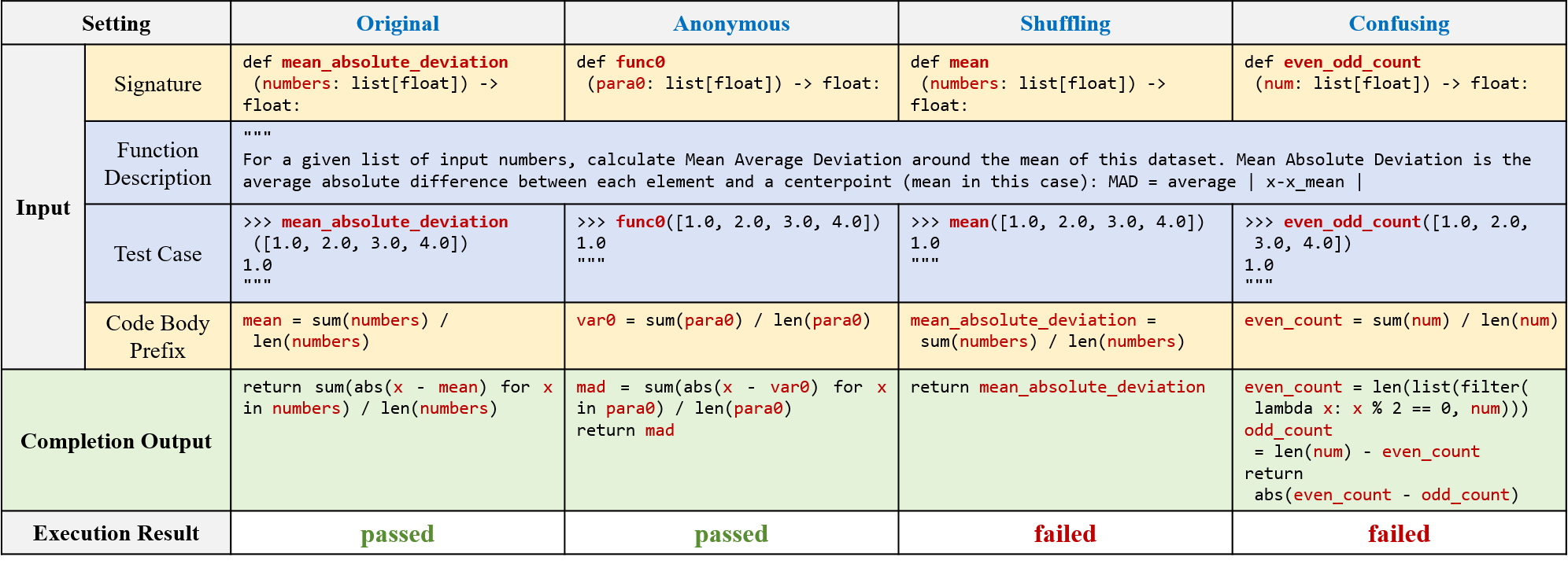}
\caption{A case study of code completion illustrating the guiding effect of lexical information. Outputs are generated by GPT-3.5-turbo.}
\label{fig:code_complete_case}
\Description{A case study of code completion illustrating the guiding effect of lexical information.}
\end{figure}

Figure~\ref{fig:code_complete_case} shows an example whose original program computes mean absolute deviation. Under the original and anonymous settings, the model produces the correct solution. After confusing renaming introduces identifiers such as \texttt{even\_odd\_count}, the model instead generates code that counts odd and even numbers and returns the absolute difference between \texttt{even\_count} and \texttt{odd\_count}. Because the natural-language HumanEval description is not provided, this output reflects a conflict between the misleading identifier cues and the operations and data flow present in the partial code, with the generated continuation following the former. In the shuffling case, swapping \texttt{mean} and \texttt{mean\_absolute\_deviation} causes the model to return the wrong variable, consistent with interpreting that variable through its new name.

In this example, the generated continuation follows the functionality suggested by the renamed identifiers despite conflicting evidence in the partial implementation. As a single case, it illustrates the pattern but does not establish its frequency.

~

Together, the selection results, summarization comparisons, and completion case provide converging evidence that misleading identifiers can directionally influence outputs in the evaluated tasks. The evidence goes beyond an undirected performance drop, while remaining observational with respect to the model's internal mechanism.

\subsection{Impact of Identifier Type}

After examining the directional influence of misleading names, we further compare the effects of modifying different identifier types. Because lexical properties are difficult to quantify directly, this analysis uses within-sample replacements grouped by identifier role.

Function names, parameters, and local variables occupy different syntactic roles and may differ in typical descriptive content. Function names often summarize overall behavior, parameters describe inputs, and local variables describe intermediate values. These are tendencies rather than fixed naming rules.

For the experiment, we randomly select one identifier in the original code snippet for modification and classify it into one of these three groups. We then statistically analyze the performance changes for each group.

\paragraph{Experiment Results}

Table~\ref{tab:onetap} reports the single-identifier results. Replacing a function name yields the lowest average ROUGE-L score, followed by parameters and local variables. Function names also have the lowest mean occurrence count (1.11), so raw frequency alone does not explain the ordering. The aggregate comparison does not, however, rule out other differences among identifier types.

\begin{table}[htbp]
\centering
\small
\caption{Results of single-identifier replacement, shown in average code summarization ROUGE-L performance per type of identifier renamed, together with the average occurrence number of each type.}\label{tab:onetap}
\begin{tabular}{cccc}
\toprule
\textbf{Identifier Type}&\textbf{Function Name} & \textbf{Parameter Name} &\textbf{Variable Name}  \\
\midrule[1pt]  
Average ROUGE-L &16.78 &18.49 &19.03 \\
Average Occurrence & 1.111& 3.255 &2.469 \\
\bottomrule
\end{tabular}
\end{table}

Identifier types also differ in syntactic position, scope, and typical descriptive content, and these factors are not separated by this experiment. We therefore interpret the table as an association between identifier role and sensitivity, not as a causal estimate of identifier type alone.

\subsection{Section Summary}

RQ2 contributes two observations. First, donor comparisons show directional influence from misleading names in the evaluated tasks. Second, single-identifier replacement yields different average sensitivities across identifier roles, with the largest decrease for function names. Together with RQ1, these findings support the lexical-overemphasis interpretation within the tested settings, without implying that all models or tasks exhibit the same degree of reliance.

\section{RQ3: Probing the Mitigability of Lexical Overemphasis}

The preceding results identify settings in which misleading or uninformative names reduce reliability. RQ3 probes the persistence of this behavior through two intervention families: (1) transformed examples or instructions at inference time and (2) fine-tuning data with reduced lexical information. These interventions test whether awareness at inference time or adaptation during training readily reduces the gap between original and lexically modified code.

\subsection{Prompt-Based Diagnostic Interventions}\label{sect:prompt-interventions}

Leveraging the instruction-following capability of larger models, we use in-context learning to test whether explicitly alerting a model to lexical changes reduces its sensitivity~\cite{zhang_transfer_2023,survey_on_cot}. Specifically, we enhance prompts with transformed examples and transformation guidelines. If lack of awareness is the main source of the observed behavior, these interventions should reduce the gap between original and confusing naming conditions.

We now elaborate on the prompt designs we employed and their underlying rationale. 

\begin{itemize}
    \item \textbf{In-context learning via perturbed examples} 
    
     Based on the few-shot learning setting in the previous task, we incorporate typical perturbed examples with the correct answer into the prompt. This intervention tests whether demonstrations of lexical variation help the LLM discount unreliable names.
     
    \item \textbf{In-context learning via transformation}
    
    Inspired by Chain-of-Thought prompting~\cite{cot}, we incorporate an intermediate reasoning step. Specifically, we inform the model that variable names may have been altered and instruct it to sanitize the input code via anonymous renaming before generating a response, with an example demonstrating the process. This tests whether an explicit sanitization step can mitigate the misleading effect.
    
\end{itemize}

\begin{figure}[htbp]
\centering
\includegraphics[width=\textwidth]{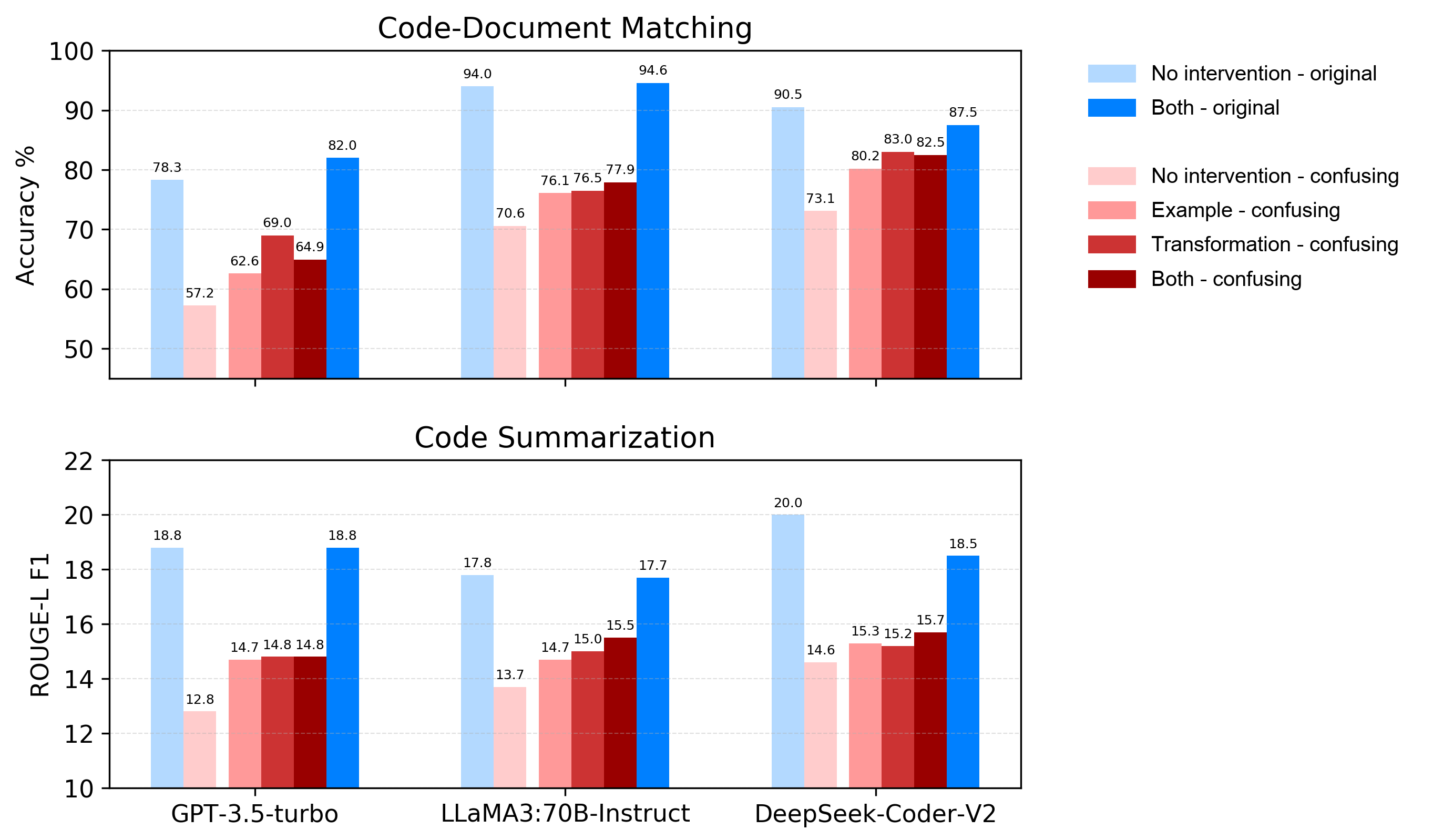}
\caption{Performance under the prompt-based diagnostic interventions. The results of each model are divided into "No intervention" on the left and "Intervention" on the right. The evaluation settings (original vs. confusing) are indicated by colors. In the legend, "Both" refers to using both interventions simultaneously.}\label{fig:defense}
\Description{Under several interventions, performance in the Confusing condition improves, but a gap relative to the Original condition remains.}
\end{figure}

Figure~\ref{fig:defense} reports the prompt interventions under Original and Confusing naming. Original-condition scores change little in these results, whereas several Confusing-condition scores improve. For GPT-3.5-turbo and DeepSeek-Coder-V2, the best observed intervention approaches the Perturbative score in Table~\ref{tab:main_result_code_summary} but remains below the Anonymous and Original scores.

These results suggest that explicit examples and transformation instructions can reduce part of the measured gap. None of the tested prompt-based interventions eliminates the gap between Original and Confusing naming.

\subsection{Fine-Tuning-Based Diagnostic Interventions}

For the smaller open-source models, we use fine-tuning rather than instruction prompting as the second diagnostic intervention. We compare the original training set with a fully anonymized version and a mixed set containing equal proportions of original and anonymized examples. We do not train directly on Confusing examples because donor selection does not define a stable target distribution. Models fine-tuned on the original data serve as the reference.

\begin{table}[htbp]
\small
\centering
\caption{Effect of fine-tuning-based diagnostic interventions on code summarization}\label{tab:finetune defense}
\begin{tabular}{llccc}
\toprule
\multirow{2}{*}{\textbf{Evaluated Model}}&\multirow{2}{*}{\textbf{Training Data}}& \multirow{2}{*}{\textbf{Original}} &\multicolumn{2}{c}{\textbf{Lexically Modified}}\\
\cline{4-5}
 & & &\textbf{Anonymous} &\textbf{Confusing}  \\
 \cline{1-5}
 \multirow{3}{*}{\textbf{CodeBERT}}&original data  &19.50&13.75&9.84\\
&anonymized data  & 18.49&15.83&13.83\\
 &mixed data (1:1 shuffled) &19.31&15.91&10.37\\

  \cline{1-5}
 \multirow{3}{*}{\textbf{CodeT5}}&original data&22.64&16.75&11.69 \\
&anonymized data  &20.85&19.24&14.95 \\
 &mixed data (1:1 shuffled)&22.68&19.28&12.34 \\
  \cline{1-5}
 \multirow{3}{*}{\textbf{DeepSeek-Coder-1.3B}}&original data  &22.22&13.63&12.31\\
&anonymized data  &20.73&17.87&15.95 \\
 &mixed data (1:1 shuffled) &21.34&17.81&13.17\\
 
 \bottomrule
\end{tabular}

\end{table}

Table~\ref{tab:finetune defense} shows that training on anonymized data yields the largest observed gains under Confusing naming for all three models, accompanied by lower Original-condition scores. Mixed-data training retains more of the Original performance and improves Anonymous performance, but produces smaller gains under Confusing naming.

Although these interventions improve some Anonymous and Confusing scores, a gap relative to Original remains for the tested models and fine-tuning settings.

\subsection{Section Summary}

We used representative mitigation strategies as diagnostic interventions. Some improve performance under lexically modified conditions, but none eliminates the measured gap. Together with RQ2, this establishes the persistence of lexical overemphasis across the evaluated prompt- and fine-tuning-based interventions. The broader scope of this conclusion is addressed in Threats to Validity.

As a secondary observation, anonymization avoids actively misleading names but also removes potentially useful information. Prior work reports comparable performance on anonymized code in some scenarios~\cite{yan_towards_2021}. Determining when anonymization is beneficial requires a separate robustness evaluation and is not established by RQ3.

\section{Discussion}

Across the evaluated models and primary tasks, the three RQs establish lexical overemphasis as a pervasive pattern that persists under the tested interventions. The type-inference control qualifies its magnitude: when the answer is locally determined without the target name, naming effects are smaller and directionally inconsistent. The larger effects in the original tasks partly reflect settings in which lexical information can be useful, while RQ2 shows that misleading names can nevertheless guide outputs. We next discuss possible contributions from input processing and training tendencies, treating them as hypotheses rather than causes established by output behavior.

\subsection{Data Processing of LLMs}

\begin{figure}[htbp]
\centering
\includegraphics[scale=0.45]{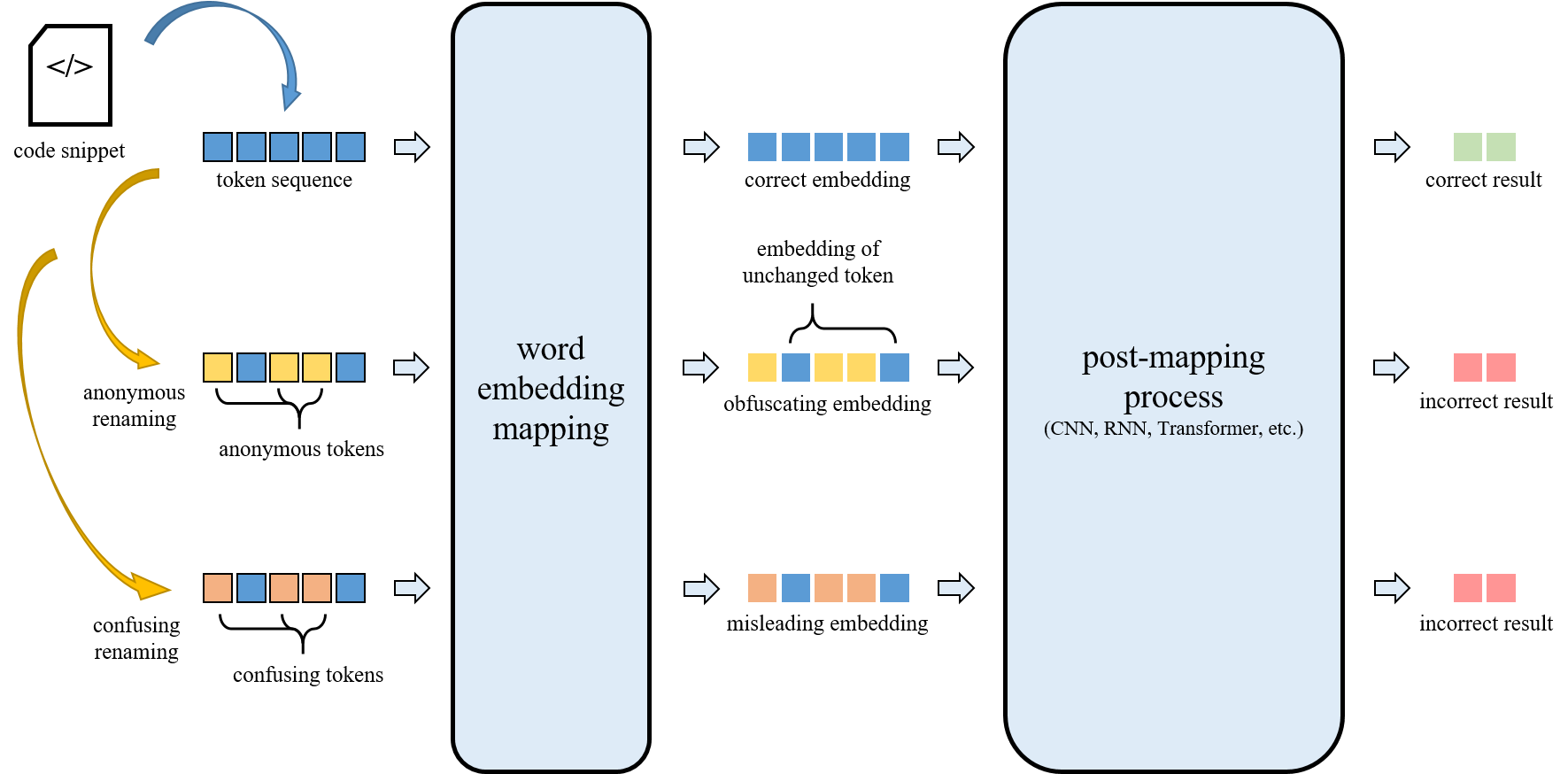}
\caption{A schematic illustration of how lexical perturbation changes an LLM's input representation and may affect its output.}\label{fig:process}
\Description{A schematic illustration of how lexical perturbation changes an LLM's input representation and may affect its output.}
\end{figure}

Face/Off changes the identifier tokens presented to a model and therefore changes its input representation, while eligible renamings preserve program behavior. This representational change need not alter a task-relevant conclusion when the unchanged implementation determines the answer. Figure~\ref{fig:process} depicts the possible pathway schematically rather than identifying a causal mechanism.

The experiments show that task-relevant conclusions are not always stable under these changes. Semantically rich identifiers may receive substantial weight during processing~\cite{yang_how_2024}, so removing or contradicting them can affect the output even when the relevant program evidence is unchanged. We interpret this as a \textbf{reliability tension}, not as a rejection of software naturalness or proof of an architectural mismatch: identifier meanings are useful predictive signals, but they are not guaranteed to be correct.

\subsection{Training Tendencies of LLMs}\label{sect:shuffling}

One possible explanation is that identifier associations remain highly predictive during training and are therefore weighted strongly at inference time. Our experiments do not determine where in the model this influence arises. To probe how much summarization performance depends on exact structural relations rather than the available token set, we conduct an exploratory shuffling experiment with GPT-3.5-turbo.

We shuffle identifiers or all tokens while retaining the token multiset. Unlike the standard Face/Off conditions, these transformations are not semantics-preserving: identifier shuffling can violate binding and syntax, and all-token shuffling destroys syntactic structure.

\begin{table}[htbp]
\centering
\caption{The ROUGE-L F1 results of code summarization on shuffled code snippets using GPT-3.5-turbo, together with reference values. Note that Face/Off shuffle is the only semantics-preserving method among the three.}\label{tab:shuffle_result}

\begin{tabular}{lcccc}
\toprule
 \multirow{2}{*}{\textbf{Shuffling Method}}&\multicolumn{4}{c}{\textbf{Proportion} (\%)} \\
 \cline{2-5}
 & 0 & 25 & 50 & 100 \\
\hline
All Tokens Shuffle & \multirow{3}{*}{18.75}& 17.97 & 17.80 & 15.04 \\
Identifiers Shuffle & & 18.20 & 17.76 & 14.87 \\
Face/Off Shuffle &  & - & - & 16.88\\

\bottomrule
\end{tabular}

\end{table}

Table~\ref{tab:shuffle_result} shows modest ROUGE-L decreases at 25\% and 50\% shuffling and larger decreases at 100\%. Even at full shuffling, the scores remain close to the Anonymous reference. This is compatible with lexical statistics supporting coarse one-sentence summaries after syntax is disrupted, but it does not show that the model internally behaves as a bag-of-words system.

This exploratory result is task-specific. Code summarization may reward topic-level lexical overlap without requiring fine-grained functional accuracy, the single-sentence prompt may amplify this effect, and identifier shuffling also disrupts syntax and binding. We did not observe a comparable pattern in our attempts on other tasks. Within these limits, full shuffling shows that lexical overlap can sustain part of the summarization score after token order is destroyed, but it neither quantifies structural reasoning nor establishes a general model mechanism.

\subsection{Section Summary}

The behavioral results identify a reliability tension between informative identifier cues and program invariance under eligible renamings. Strong lexical associations learned during training are one plausible contributor, and the shuffling result is compatible with coarse lexical statistics supporting summarization. Neither observation directly identifies an internal mechanism.

\section{Threats to Validity}

\subsection{Internal Validity}

For the standard Face/Off conditions, we consistently rename eligible local bindings while retaining the program structure and operational behavior, which is designed to isolate identifier information. Residual threats include transformation errors and task-specific interactions between code and text; we address these through implementation checks and the documentation procedure in Section~\ref{sect:method}. In the type-inference control, parse, compile, binding-occurrence, and alpha-equivalence checks reduce transformation risk, and paired tests are clustered by project because multiple targets can come from the same project. However, target eligibility was established through conservative automatic analysis rather than an independent human identifier-independence review. We therefore present this experiment as a supplementary semantic control rather than a confirmatory estimate for a broader type-inference population. The extended external-identifier experiment is less controlled because it inserts a binding block. Its dead-code condition estimates part of the insertion effect but does not isolate decontextualization perfectly. Table~\ref{tab:detailed match task} shows the same condition ordering across the tested prompts, but other prompts could yield different magnitudes or patterns. Finally, benchmark contamination~\cite{magar2022data,golchin2023time} could affect both baseline performance and measured sensitivity, so the results should not be interpreted as contamination-free estimates.

\subsection{Construct Validity}

The naming conditions and identifier roles operationalize lexical-information quality, but we lack a direct quantitative measure that captures every aspect of informativeness, naturalness, or ambiguity. The observed patterns are therefore conditional on the evaluated tasks, models, programs, and transformations.

A potential concern is that tasks such as code summarization couple identifier words with ground-truth text through lexical-overlap metrics. For the CodeSearchNet-based tasks, explicit references to the subject function follow the corresponding code condition, while parameter words in natural-language prose are retained to preserve the description's semantic content (Section~\ref{sect:method}). Exact subject-function names occur in only 2 of the 600 CodeSearchNet descriptions used in this evaluation. Consequently, condition-specific synchronization changes the accompanying text for only these cases; parameter names and randomly selected negative descriptions remain unchanged. This low incidence limits the extent to which the aggregate performance patterns can be attributed to reference-text rewriting. The targeted procedure does not make every identifier-like word lexically aligned. HumanEval does not present its natural-language description to the model. The directional analyses in Section~\ref{sect:guiding} provide complementary evidence: donor-reference similarities exceed random-pair comparisons, which is less consistent with arbitrary label disruption than with influence from donor identifiers. These analyses reduce, but do not eliminate, the construct-validity concern because the tasks and metrics still differ in their lexical requirements.

Finally, the mechanism-level discussion is based on output behavior and exploratory perturbations rather than direct access to internal causal processes. The proposed explanations should therefore be treated as hypotheses for future mechanistic and experimental work.

\subsection{External Validity}

We evaluated models with different architectures, sizes, and access types across four code-comprehension tasks. Repeated patterns across these settings provide within-study replication, but they do not guarantee generalization to other models, languages, tasks, or repositories. In particular, the HumanEval code-completion problems are relatively small, and the controlled snippets do not capture dependencies and workflows found in large codebases. The type-inference control deliberately selects cases with locally sufficient evidence, uses a closed set of 15 labels, and is dominated by primitive types; this conservative construction improves isolation but makes the task easier and limits generalization to user-defined, repository-dependent, or unrestricted types. Its model set also differs from the original four-task evaluation, so we do not use it for direct numerical comparison. Future work should test the same hypotheses in repository-level, documentation-rich, tool-augmented, and more difficult type-inference settings.

\subsection{Ecological Validity}

Confusing names and systematic placeholders are controlled stress-test conditions, not estimates of how frequently the same patterns occur in everyday development. Anonymous naming resembles some information-depleted artifacts, such as minified, obfuscated, decompiled, or poorly named code, but it is not a general proxy for all low-quality code. Consequently, our results establish sensitivity under the tested interventions; they do not estimate the population-level failure rate in real projects.

Human program comprehension is also multi-source and task-dependent. Descriptive identifiers and recognizable beacons can help programmers, while misleading beacons can also produce false initial interpretations~\cite{schankin2018identifiers,wiedenbeck1991initial}. Evidence that experts form representations from procedural relations~\cite{pennington1987stimulus} does not imply that humans reason independently of names. Moreover, professional developers use browsers, documentation, and tools during comprehension~\cite{xia2018measuring}. Our experiment intentionally restricts the context supplied to the model and, where a task pairs code with documentation, controls that textual channel together with the naming condition. This restriction improves internal validity by isolating identifier information, but reduces ecological realism. It is a property of the experimental design, not evidence that an LLM cannot use documentation or tool feedback. We therefore compare behavior only under the supplied context and do not claim a categorical difference between human and model comprehension.

\section{Conclusion and Implications}

This empirical study investigates lexical overemphasis in LLM code comprehension, characterizes its behavior under controlled identifier transformations, and uses diagnostic mitigation attempts to evaluate its persistence. Within the Face/Off framework, lexical overemphasis is pervasive across the evaluated models and original four tasks, and persists under the representative prompt- and fine-tuning-based interventions we test. A targeted type-inference control confirms a boundary on this pattern: in its conservatively screened subset, naming effects are smaller and directionally inconsistent when local program evidence suffices to determine the answer. Together, the results identify a task-dependent reliability tension between useful distributional naming cues and task-relevant structural evidence; they neither reject software naturalness nor imply that identifiers should be ignored.

For practitioners, the results motivate caution when identifier quality is uncertain and independent validation when correctness is consequential; they do not establish anonymization as a universally beneficial safeguard. For researchers, they motivate training and evaluation methods that retain the predictive benefits of natural code while checking whether conclusions remain stable under semantics-preserving changes and grounded in program structure.

\bibliographystyle{ACM-Reference-Format}
\bibliography{ref}

\end{document}